\documentstyle[12pt,epsf,epsfig,wrapfig]{article}
\newcommand{\itemnew}{\item \vspace*{-3mm}}
\begin{document}

\begin{center}
{\bf
SPIN PHYSICS PROGRAM IN THE U70 POLARIZED PROTON BEAM \\
}\
\vskip 5mm
V.V.Abramov$^{1}$,
S.I.Alekhin$^{1}$,
A.S.Belov$^{2}$,
V.I.Garkusha$^{1}$,
A.V.Efremov$^{3}$,
P.F.Ermolov$^{4}$,
S.V.Ivanov$^{1}$,
V.I.Kravtsov$^{1}$,
V.I.Kryshkin$^{1}$,
A.V.Kubarovsky$^{4}$,
A.K.Likhoded$^{1}$,
V.V.Mochalov$^{1}$,
D.A.Morozov$^{1}$,
L.V.Nogach$^{1}$,
S.B.Nurushev$^{1}$,
A.F.Prudkoglyad$^{1}$,
V.N.Ryadovikov$^{1}$,
I.A.Savin$^{3}$,
P.A.Semenov$^{1}$,
Y.M.Shatunov$^{5}$,
S.R.Slabospitsky$^{1}$,
L.A.Tikhonova$^{4}$,
D.K.Toporkov$^{5}$,
S.M.Troshin$^{1}$,
E.F.Troyanov$^{1}$,
M.N.Ukhanov$^{1}$,
\underline{A.N.Vasiliev}$^{1\dag}$

\vskip 5mm
{\small
(1){\it
Institute for High Energy Physics, Protvino, Russia
}\\
(2){\it
Institute of Nuclear Research, Troitsk, Russia
}\\
(3){\it
Joint Institute of Nuclear Research, Dubna
}\\
(4){\it
D.V. Skobeltsyn Institute of Nuclear Physics, Lomonosov Moscow State
University, Moscow, Russia
}\\
(5){\it
Budker Institute of Nuclear Physics, Novosibirsk, Russia
}\\
$\dag$ {\it
Email : Alexander.Vasiliev@ihep.ru \\
}}
\end{center}

\vskip 5mm
\begin{abstract}
A possibility to accelerate a high intensity polarized proton beam
up to 70 GeV at the IHEP accelerator,
extract it from the main ring and deliver to several experimental
setups is being studied now. We propose to study a wealth of
single- and double-spin observables in various reactions using
longitudinally and transverserly polarized proton beams at U70.
The proposed measurements can be done at the existing detectors as well
as require to create a few new experimental setups at U70.
\end{abstract}

\vskip 8mm

\section * {Introduction}
\vskip 5mm

We propose to produce polarized proton beam from the polarized
atomic beam source,
accelerate it in the 1.5 GeV booster and then in the
U-70 main ring up to 70 GeV, extract it from the main ring and deliver
to several experimental setups to:
\begin{itemize}

\itemnew  measure the gluon and quark polarization in longitudinally polarized
protons in charmonium production to help to solve the problem of
``spin crisis'';

\itemnew measure
the double-spin transversity distribution in transversely polarized
protons by using the Drell-Yan muon pair production;

\itemnew measure dependence of single-spin asymmetries(SSA) on
separate kinematic variables p$_T$ and x$_F$ and hadron type to study the
possible non-perturbative origion of SSA;

\itemnew measure miscellaneous spin parameters in hyperon
production at moderate transverse momenta to learn about role of strange
quarks in the spin structure of nucleon;

\itemnew measure polarization and spin correlation parameters in elastic
pp-scatering in the hard scattering region in order to check the
QCD predictions.

\end{itemize}

Although years of experimental effort at miscellaneous accelerators
have provided a lot of information about the QCD hard scattering
and the parton structure of the proton, there is no corresponding body of data
on the spin-dependence of the elementary interactions and the spin structure
of the proton. High intensity polarized proton beam accelerated up to 70
GeV and extracted from the U70 main ring offers the opportunity to study
the unique properties of the spin variable at large x to increase the
understanding of these fundamental quantities.

%The  spin studies program have several well defined goals and among them:
Final goals of spin physics at U70 are :
\begin{itemize}
\itemnew
study the spin structure of the proton, i.e., how the proton's spin
state can be obtained from a superposition of Fock states with different
numbers of constituents with nonzero spin;
\itemnew
study how the dynamics of constituent interactions depends on  spin
degrees of freedom and on the flavors;
\itemnew
understand chiral symmetry breaking and helicity non-conservation on the
quark and hadron levels;
\itemnew study the overall nucleon spin structure in the range of
 moderate p$_T$(up to 5 GeV/c), and the long range QCD dynamics (confinement),
including study non-perturbative interactions of massive constituent
quarks with an effective color field of flux tubes, produced by confined
quarks and gluons.
\end{itemize}

 These issues are closely interrelated at the hadron level and
 the results of the experimental measurements
 are to be interpreted in terms of hadron  spin
structure convoluted with the constituent interaction dynamics.

\section {Acceleration of Polarized Protons at U70}
\vskip 5mm

To do this spin physics, the polarized proton beam with an intensity
up to  10$^{12}$ protons per spill and energy up to 70 GeV needs to be
accelerated. This process starts from an
intensive polarized proton source. A polarization value of the beam
needs to be measured continuosly by polarimeters. The setups
should include polarized targets if double-spin measurements are
being planned.\\

To achieve high luminosity in the interactions of polarized beams
with the targets, the intensity of the planned polarized proton
source should be sufficient. This source is expected to be designed
and built at
the Institute of Nuclear Research, Troitsk. It will be an atomic
beam-type polarized ion source. The source might produce up to 5 mA
of H$^-$ with 90$\%$ polarization. \\

During acceleration, the polarization may be lost when the spin precession
frequency passes through a so-called depolarizing resonance. These resonances
occur when the spin tune
$\gamma$$\times$G (where G=1.793 is the anomalous magnetic moment of
the proton and
$\gamma$=E/m) is equal to an integer number (imperfection resonances), or
equal to kP$\pm$$\nu_z$ (intrinsic resonances). Here P=12 is the superperiod
of the U70 accelerator, $\nu_z$=9.9 is the vertical betatron tune, and k is
an integer. Imperfection resonances are due to vertical closed orbit errors
and intrinsic resonances are due to the vertical betatron motion.
It is possible to preserve a high polarization value, sort of 70$\%$
over the acceleration
of the polarized proton beam up to 70 GeV if one installs three partial
siberian snakes in the main ring of the U-70 with the snake ``strength''
W=$\phi$ /2 $\pi$ = 0.18. Three definite long streight line
sections (4.87 m each) in the U-70 main ring
need to be cleaned up from the existing equipment.  Much more details
can be found in the report of  Yu.M.Shatunov \cite{spin2005_shat}.\\
%Several milestones will have to be achieved :
%\begin{itemize}
%\itemnew
%12 superconducting helicoidal magnets (three snakes) with the length
%of 0.6-0.7m and magnetic field 5-6 T need to be designed and built ;
%\itemnew
%emittance of the U70 beam will have to be preserved at the level
%of 10-15 mm mrad;
%\itemnew
%all the magnets in the U-70 will have to be surveyed in the horizontal
%plane within +/-5 mm accuracy;
%\itemnew
%betatron tunes should be as follows : $\nu_z=9.9$  and
%$\nu_x=9.7$.  \\
%\end{itemize}

There should be two types of polarimeters at U70 : 1) absolute polarimeter
to determine a value and a sign of polarization with a high accuracy and
2) less accurate but fast relative polarimeter. The both polarimeters
should be based on Coulomb-Nuclear Interference (CNI) effect. The absolute
polarimeter could be built with the use of a polarized jet target, while
the relative polarimeter with the use of Carbon thin target. The absolute
polarimeter calibrates the relative one. The polarization of proton beam
circulated in the U70 averaged over 100 hours data taking could be measured
to 10$\%$ statistical accuracy with the use of hydrogen jet.
The detail description of the polarimetry for U70 can be found in the report
of S.B.Nurushev \cite{spin2005_nur}.\\

Longitudinally and transversely polarized targets are needed for
double-spin measurements. The effect
of a high intensity polarized proton beam is to deposit a significant amount
of heat in the target. It is thus necessary to use $^4$He evaporation
refrigirators to cool the target, instead of dilution refrigirators, even
though the lower temperatures achievable in dilution refrigirators often
results in higher polarizations. A second effect of the high intensity
beam is a significant amount of
radiation damage in the target material. Materials that are chemically
doped typically perform poorly under conditions of high-radiation
damage, so radiation-doped materials, typically ammonia
($^{15}$NH$_3$) or lithium ($^6$LiD)
hybrides, are used. These continuous pumping polarized
targets are the best ones to work with
high intensity beams up to 10$^{11}$ p/[cm$^2\times$ s].

\section{Single-Spin Asymmetries in Inclusive Processes}

The studies of the spin effects in
inclusive processes probe the  spin dependence of
the incoherent hadronic interaction dynamics.
The cross--sections of the hard production processes
are described in the
perturbative QCD as a convolution integral of parton
cross-sections with the light--cone parton densities.
 The primary goal of the single-spin measurements
 with hadronic final states
 would be a study where the onset of perturbative QCD
regime occurs. It is usually assumed, that single-spin transverse
asymmetries in inclusive process
$A+B\to C+X$, where A is a polarized hadron, have higher twist origin
\cite{kane,efremt}.
The contribution of higher twists should be small at high energies
and at small distances $l\sim 1/Q$.
There are some indications that such contributions are small even at
not too high energies and $Q^2$ values.
In particular, it follows
from the recent data on the spin structure function
$g_2(x)$ obtained at SLAC. If it is the case, the observed
significant one-spin asymmetries in hadronic processes are to be
associated with the
manifestation of a nonperturbative dynamics.

The measurements of one-spin transverse
asymmetries will be important probe of the
 chiral structure of the effective QCD Lagrangian.

The available experimental data
are at some variance with PQCD predictions: these data
do not at least show up
tendency to converge to the vanishing single-spin asymmetries
in inclusive and elastic hadron productions.
Several mechanisms have been proposed for the explanation
of the observed single-spin effects.

%\begin{wrapfigure}{R}{6.2cm}
\begin{wrapfigure}{R}{8.5cm}
%\mbox{\epsfig{figure=fods.eps,width=5.8cm,height=5.2cm}}
\mbox{\epsfig{figure=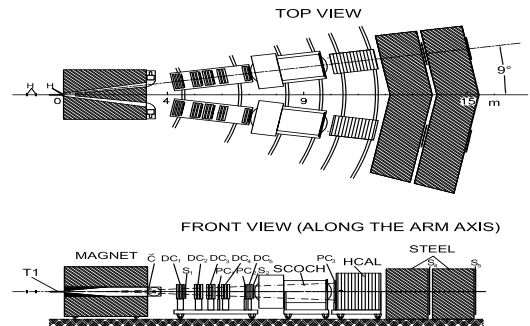,width=8cm,height=7cm}}
\caption{Experimental setup FODS.}
\label{vasiliev_fig1}
\end{wrapfigure}

Single transverse spin asymmetries in hard processes are expected
to vanish in perturbative QCD at the leading twist (twist-2) level,
however this is no longer true if one considers higher twist
contributions. Twist-3 contribution \cite{efremt} leads to a
potentially larger
asymmetry. A measurement of higher twist effects not only provides
a valuable test on perturbative QCD, it also yields information on the
hadron structure. The higher twist distributions represent correlations
between quarks and gluons inside a hadron. These correlations have specific
structures in QCD. If extracted from experimental data, they provide
useful constraints on nucleon models, in particular, the non-valence degrees
of freedom that include sea quarks and gluons.

Other world wide known possible  sources of the observed one-spin
asymmetries could be: correlation of $k_\perp$ and spin in initial state
\cite{sivs} (Sivers effect)  and fragmentation \cite{cols} functions
(Collins effect). Usually we have combinations of these effects together
with Twist-3 effect in a particular experiment and special technics are
required to extract them from observed asymmetries. There are also
other theoretical approaches to explain large single-spin asymmetry
in inclusive hadron reactions such as
 rotation of valence quarks inside a hadron \cite{boros} and the
coherent rotation of the quark matter inside the constituent
quarks \cite{asy}. The Stern-Gerlach type of force \cite{ryskin}
which acts on chromomagnetic quark moment in color flux tube field
\cite{migdal}, is able to explain many features of SSA dependence on
hadron type and kinematic variables.

The study of charged hadron inclusive reactions  at U-70 energies is
important as a clear
test of perturbative QCD regime and nonperturbative dynamical models.
In general, these studies are important for the understanding
of the QCD vaccuum and transitions between the perturbative
and nonperturbative phases. A detailed high precision study of SSA
dependence on kinematic variables and hadron type allows to discriminate
some of the proposed models and the origion mechanisms. \\

Asymmetries in inclusive production of charged pions, kaons, protons
and antiprotons can be measured by FOcusing Double Arm Spectrometer
(FODS) which is placed at the beam line 22 at IHEP.
The experimental setup FODS is shown in Fig.~\ref{vasiliev_fig1}.
It consists of the analyzing magnet, drift chambers, spectrometer
of Cherenkov radiation (SCOCH) for particle identification (charged
pions, kaons, protons and anti-protons), scintillation counters and
hadron calorimeters to make trigger on high energy hadrons. There
are two arms which can be rotated around the target center situated
in front of the magnet to change secondary particle angle.
The radiation shielding in the beam line allows to work with a
beam intensity up to 10$^{10}$ p/s, and the setup up to 10$^9$ p/s.\\

The following single-spin measurements can be done at FODS:

\begin{itemize}

\itemnew Precise measurements of $A_N$ in inclusive production of
charged pions, kaons, protons and antiprotons at hydrogen and nuclear
targets. Large $x_T$ can be achieved. To separate $p_T$ and $x_F$
asymmetry dependences, the measurements at several angles are needed
(might be done in the range of 10-130$^0$ in c.m.).

\itemnew Measurements of $A_N$ in symmetric hadron production. Symmetric pairs
($\pi$$^+$ and $\pi$$^-$,etc) are the hadrons produced in the c.m. with
about the same momenta and moving in the opposite directions. For these
processes $k_T$ is almost zero and there is no Sivers effect.

\itemnew  Measurements of $A_N$ in Drell-Yan muon pairs. There is no
fragmentation in this process - this means that there is no Collins effect.
(Additional absorbers in each arm will be installed).

\end{itemize}

\section{Measurements of Quark Transversity Distributions in a
Polarized Proton}

From deep inelastic scattering, one can measure $f_1(x)$ related to
the spin-averaged longitudinal momentum distribution of quarks in the
nucleon and $g_1(x)$ related to the helicity distribution in a longitudinally-
polarized nucleon. In addition to these two well-known structure functions,
there exists a third fundamental function, $h_1(x)$, which is a leading twist
(twist-2) distribution function like $f_1(x)$ and $g_1(x)$. This third
function, which has never been experimentally determined, is accessible
by measuring the double transverse spin asymmetry A$_{TT}$ in certain
processes with both beam and target protons transversely polarized.

A measurement of $h_1(x)$ can shed interesting light on the spin
structure of the nucleon. In fact, in non-relativistic quark models,
the transversity distribution is identical to the quark helicity distribution
$g_1(x)$. Thus, a comparison between the sizes of $h_1(x)$ and $g_1(x)$
would measure the success of these models in treating the spin degrees
 of freedom. The function $h_1(x)$ , unlike the helicity function
$g_1(x)$, cannot be measured in deep inelastic scattering due to its
different properties under chiral transformations.

Quark transversity is a new observable for understanding
the hadron
wave function in terms of bare quarks. Gluons give no contribution
to the transverse spin of the proton. It is promising to explore this
new spin observable and compare it with the longitudinal spin densities.
With transversely polarized beams the new field of transverse spin
effects can be explored in the Drell-Yan muon-pair channel by measuring
the correlation of the plane of the muon-pair to the spin axis.
This provides a clean approach to quark transversity, h$_1$(x).

A beam dump experiment might be an appropriate one to measure
transversity in Drell-Yan muon pairs. The scheme of the experiment is as
follows. A polarized beam interacts with a polarized target. There is
an absorber just after the target and finally a muon detector to
detect Drell-Yan muon pairs. An issue will be a hadronic background which
can originate from decays of charged pions and kaons before reaching
the absorber and from hadrons penetrating the material (punch-through).

\begin{wrapfigure}{R}{10cm}
\mbox{\epsfig{figure=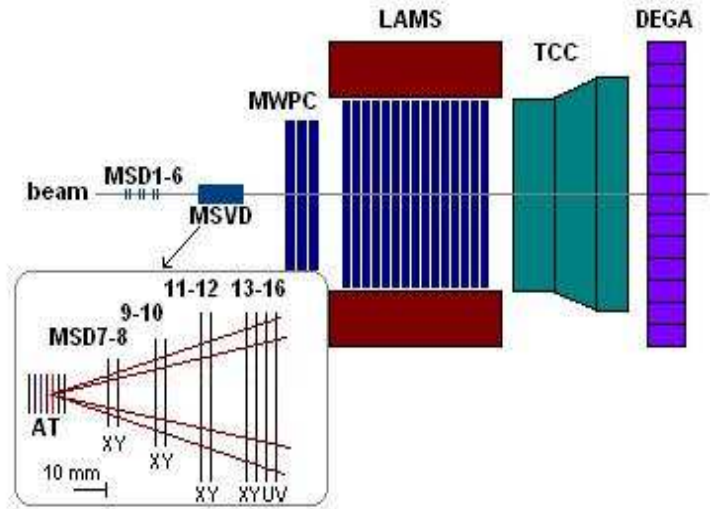,width=10cm,height=7cm}}
\caption{Experimental setup SVD-2.}
\label{vasiliev_fig2}
\end{wrapfigure}

If the polarized target can accept an intensity of 10$^{11}$ p/spill
(we expect spill duration of 3 s out of 9 s full cycle),
then the luminosity of this experiment would be about
10$^{34}$ cm$^{-2}$ s$^{-1}$ that
is three orders of magnitude higher than today at RHIC or at the planned
PAX at GSI. The region of x$_1$x$_2$=0.02-0.09 for the Drell-Yan masses
between 1.5 and 3 GeV/c$^2$ will be covered. These measurements will be
complementary to the RHIC and PAX ones, where the covered regions are
expected to be 0.004-0.02 and 0.07-0.3, correspondingly. The estimated
numbers of Drell-Yan events for 30 days at beam are about 180,000
and 24,000 for the masses of 1.5 and 2.0 GeV/c$^2$, respectively.
The A$_{TT}$ errors for these masses are expected to be in the
range of (2-4)$\%$.

%\section{Strangeness in the Hadrons}
\section{Measurements of Spin Effects in Strange Hadron Production}
It is  evident from deep--inelastic
scattering data that
strange quarks as well as gluons could play essential role in
the spin structure of nucleon.
 DIS data show that strange quarks are
 negatively polarized in polarized
nucleon, $\Delta s\simeq -0.1$.
Elastic $\nu p$-scattering data provide the value
 $\Delta s=-0.15\pm 0.08$ \cite{nu}.
The presence and polarization of strange quarks inside a hadron should
give an experimental signal in hadronic reactions also.

\begin{wrapfigure}{R}{6.2cm}
%\mbox{\epsfig{figure=lamda_mass.eps,width=5.8cm,height=5.2cm}}
\mbox{\epsfig{figure=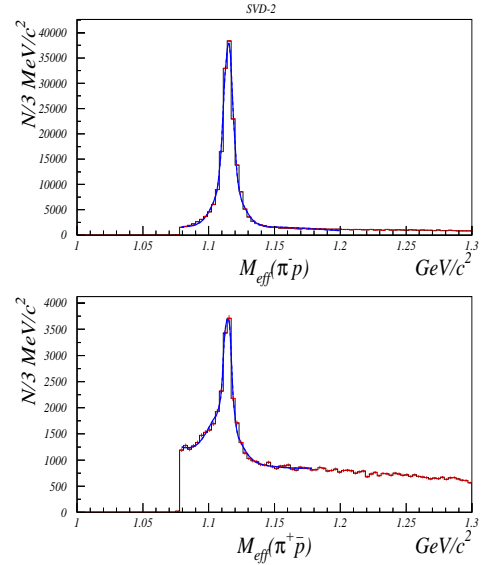,width=7.0cm,height=8.2cm}}
\caption{Spectra of $\pi^-$ p  (upper part) and
$\pi^+$ $\bar{p}$ (lower part) invariant masses.}
\label{vasiliev_fig3}
\end{wrapfigure}

Experimental
 situation with hyperon polarization is widely known and stable
for a long time.
Polarization(P$_n$) of $\Lambda$--hyperon produced in the
unpolarized inclusive $pp$--interactions is negative and energy
independent. It increases linearly with $x_F$ at large transverse momenta
($p_\perp\geq 1$ GeV/c),
and for such
values of transverse momenta   is almost
$p_\perp$-independent.

We consider the production of hyperons in the kinematic region of p$_T$
from 1 to 5 GeV/c and at large x$_F$. A systematic study of many spin
observables (P$_n$, A$_n$, D$_{nn}$, A$_{LL}$, etc.) with polarized beam and
polarized target for the different hyperon productions could be done
with the use of the U-70 polarized beam.
  The measurements can be done at the existing experimental setup SVD-2
\cite{svd_setup}(see  Fig.~\ref{vasiliev_fig2}) which is placed
at the beam
line 22 at IHEP and consists of a) the high-precision microstrip
vertex detector(MSVD) with active(Si) and passive(C,Pb) nuclear
targets(AT), b) the large aperture magnetic spectrometer(LAMS)
with two sets of MWPC(multiwire proportional chambers),
c) the multicell threshold Cherenkov counter(TCC), and d) the gamma
quanta detector (DEGA). The mass spectra of $\Lambda$-hyperons
are shown in Fig.~\ref{vasiliev_fig3} to demonstrate the setup
capability to
detect hyperons.
More than 200,000 $\Lambda$-hyperons were recently detected over one month
of data taking with a beam intensity of 0.5$\times$10$^6$ protons/s.
$\Lambda$-hyperons are very well detected in the beam fragmentation
region where large spin effects are expected.

Spin effects in strange hadron production shed light on the role of
strange quarks in the spin structure of nucleon and non-perturbative
QCD dynamics.

\section { Double Spin Asymmetry in Charmonium Production }
\vskip 5mm

The study of spin effects in some processes would yield
information on the contribution of the spin of quarks
$\Delta\Sigma$ and gluons $\Delta$G and orbital angular momenta of
quarks $\L_q$ and gluons $\L_g$ into the  hadron helicity:

\begin{equation}
1/2=1/2\Delta\Sigma+L_q+\Delta G+L_g
%\Gamma+L_g
\end{equation}

In the above sum  all  terms have clear physical interpretation,
however  besides the first one, they are gauge
 and frame dependent. Transparent discussion of the theoretical aspects
of this  sum rule and a new gauge independent one are given in
 \cite{jin}. Gluon contribution into the proton spin was worldwide studied
at HERMES, COMPASS, RHIC, JLaB and SLAC, however new data especially at
large x are very appreciated.

\begin{wrapfigure}{R}{8.5cm}
\mbox{\epsfig{figure=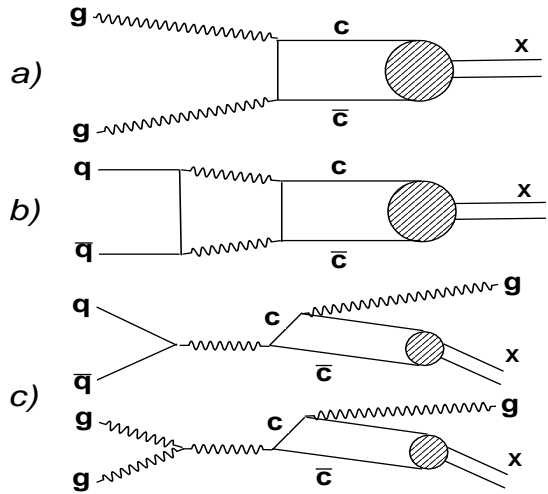,width=8cm,height=7cm}}
\caption{ Lowest-order parton fusion diagrams for hadronic $\chi$ production:
a) gluon fusion, b) quark-antiquark annihilation, c) color evaporation.}
\label{vasiliev_fig4}
\end{wrapfigure}

We propose to simultaneously measure the double-spin asymmetry A$_{LL}$
for inclusive $\chi_2$, $\chi_1$ and J/$\psi$ by utilizing the 70 GeV/c
longitudinally polarized-proton beam on a longitudinally polarized target.
Our goal is to obtain besides the quark-spin information also
the gluon-spin information from these three
processes in order to determine what portion of the proton spin is
carried by gluons. We anticipate obtaining significant numbers
of  $\chi_2$, $\chi_1$ and J/$\psi$ events. The statistical errors on
A$_{LL}$ will be small enough for the possible determination of
the spin-dependent gluon structure function in a specific x range
where the gluon polarization is expected to be sizeable. This would
be the world's first measurement of gluon-spin information and of
spin effects in charmed-particle production in hadron-hadron
interactions. The same proposal at Fermilab \cite{p838} at 200 GeV/c was not
approved in 1991. The Fermilab PAC committee did not believe that
proper number of charmonium events would be collected by using
low intensity polarized proton beam obtained from $\Lambda$-hyperon
decays.

The hadronic production of the $\chi$ states involves three parton fusion
diagrams (see Fig.~\ref{vasiliev_fig4}): gluon fusion, light quark
annihilation,  and color evaporation \cite{charmonium}.
However, the relative contributions of each subprocess and even the total
cross section for charmonium production have proven difficult to calculate
reliably. There are fairly definite predictions
for the relative production rates of the $\chi$ states which may help
distinguish among the models. The theoretical predictions for the
ratios $\sigma_1$/$\sigma_2$=$\sigma$($\chi$(3510))/$\sigma$($\chi$(3555))
are as follows \cite{ratios} : zero for gluon fusion, 4.0 for light quark
fusion and 0.6 for color evaporation. One of possible way to measure
such parton's polarization is a study
of $\chi_c$-meson production with the following decay into $J/\psi$ and a
photon and then $J/\psi \to \ell^+ \ell^-$. It was shown
in~\cite{jpsipol}, that
the angular distribution of the final photon and lepton pairs provides a
direct way to measure the polarization of the initial quarks and gluons.

In this experiment the separation
of $\chi_1$(3510 MeV) and $\chi_2$(3555 MeV) is possible.
The matrix element for $\chi_1$ production via gluon fusion is calculated
to be zero according to the lowest-order QCD \cite{charmonium}. If few
$\chi_1$ events are detected, then gluon fusion is dominant in
$\chi_2$ production. If the number of $\chi_1$ events is significant, we need
to also include other processes in the calculations to determine
$\Delta$G/G. Note that A$_{LL}$(p$\uparrow$N$\uparrow$ $\to$ $\chi_1$ + X)
and the $\chi_1$ $\to$ J/$\Psi$ + $\gamma$ decay angular distribution
will be measured simultaneously in this case, providing an additional
input for understanding the production process and the value of
$\Delta$G/G near x=0.3. We propose to measure the A$_{LL}$ asymmetry
in the J/$\Psi$ production via the J/$\Psi$ $\to$ e$^+$e$^-$ channel.

If all three charmonium production processes contribute, the measurement
of $\chi_2$, $\chi_1$, and J/$\Psi$ become equally important. The
A$_{LL}$ ( $\chi_2$, $\chi_1$, J/$\Psi$) provide a test to various models,
which predict opposite A$_{LL}$ signs. The signs and magnitudes
of A$_{LL}$($\chi_2$) and  A$_{LL}$(J/$\Psi$)
will provide crucial information on the production mechanism(s), if
the $\chi_1$ production at 70 GeV in pp-interactions is not negligible
compare to the $\chi_2$ production. A large value of A$_{LL}$ will indicate
a sizeable $\Delta$G/G independent of models. \\

%\begin{wrapfigure}{R}{8.5cm}
%\mbox{\epsfig{figure=setup.eps,width=8cm,height=7cm}}
%\caption{Schematic view of experimental setup to measure A$_{LL}$ in
%$\chi$ and J/$\Psi$ production.}
%\label{fig:setup}
%\end{wrapfigure}

The experimental setup in the open geometry configuration
%(see Fig.~\ref{fig:setup})
will consist of electromagnetic
calorimeter, proportional chambers and plastic scintillator-pad detector
for the charmonium trigger.

In order to separate $\chi_1$(3510 MeV) and $\chi_2$(3555 MeV) peaks,
the energy resolution of the calorimeter is important, especially for
the produced $\gamma$'s. According to the decay kinematics of $\chi_2$,
the $\gamma$'s are effectively detected at very forward angle (up to 100 mrad).
A calorimeter for detection of these $\gamma$'s must have good energy
resolution with fast response to handle the rates.

The central part ($\Theta_{lab}$ from 10 mrad to 100 mrad) of the calorimeter
system consists of 1152 blocks of lead tungstate
(2.8 $\times$ 2.8 $\times$ 22 cm$^3$ per each block) to ensure good energy
resoltion of the $\gamma$ detection. This is an array of 34x34 blocks
with a hole of 2x2 blocks in the center for non-interacted beam.
The properties of lead tungstate (PWO)
calorimeters have been extensively studied at IHEP over last several
years \cite{nim8}.  The energy resolution of ~2$\%$ at
E = 1 GeV has been measured. The 1875 lead-glass counters
(3.81 $\times$ 3.81 $\times$ 45 cm$^3$ per each block)
cover a large area ($\Theta_{lab}$ from 100 mrad to 200 mrad).
This is an array of 50x50 blocks with a hole of 25x25 in the center
to accept the lead tungstate blocks.

The GAMS-2000 experiment \cite{gams} at IHEP used the same kind of
lead-glass blocks with a similar setup in this proposal. The $\chi$ states
have been detected by measuring e$^+$,e$^-$, and $\gamma$. Clear J/$\Psi$ and
$\chi$ peaks were observed with an open geometry configuration at a beam energy
of 40 GeV, but no separation between the two $\chi$-states was
performed due to poor energy resolution.

The proportional chambers placed between the target and the calorimeter
serve to track e$^+$ and e$^-$ particles and assure that there are no charged
tracks in the $\gamma$ direction.
The scintillator-pad trigger hodoscope containing 100 pads has
a segmented mosaic structure. Our first estimate shows us that
we might anticipate
about 10,000 reconstructed $\chi_2$ events/month and more than 50,000
J/$\Psi$ events/month. A new experimental setup will have to be built
to accomplish these measurements.\\

\section { Polarization in Elastic Scattering }
\vskip 5mm
In perturbative QCD, there are several
mechanisms that could give important contributions to
fixed-angle elastic scattering. However, due to small cross--section
studies of this exclusive process could be carried out in the
region where nonperturbative effects are essential.
There are  several models based on nonperturbative dynamics
predicting
significant nonzero analyzing power at fixed angles(for example,
\cite{mqm,diq}).\\

The measurements of polarization in elastic pp-scattering (parameter A$_N$)
can be done at the experimental setup ``SPIN at U70'' which is placed
at the beam line 8 at IHEP. Both particles, forward and recoil
protons will be detected
by scintillation hodoscopes (forward arm) and by drift chambers (recoil arm).
A resolution of the drift chambers will be about 200 $\mu$m.
The setup can afford an intensity of the polarized proton beam up to
10$^{12}$ p/spill. Particle identification will be performed by Cherenkov
counters in the both arms and additionally by a time-of-flight
technique in the recoil arm.

The parameter A$_N$ in elastic pp-scattering will be measured
at 70 GeV in a wide p$_T$$^2$ region - from 1 to 12 (GeV/c)$^2$.
For 200 hours of data taking statistical errors in A$_N$ will be less
than 1$\%$ for p$_T$$^2$ up to 6 (GeV/c)$^2$. For 600 hours
errors in A$_N$ will be about 3$\%$ for 10 (GeV/c)$^2$ and about 6$\%$
for 12 (GeV/c)$^2$.
 The results will be used to discriminate
among several models describing the polarization in elastic
scattering in the hard interaction region.

\newpage
\section * { Conclusion }
\vskip 5mm

To accelerate the polarized proton beam in the existing U70 accelerator
up to 70 GeV with intensity up to 10$^{12}$ protons/spill and
polarization up to 70$\%$, the following main tasks need to be completed:

\begin{itemize}

\itemnew a polarized H$^-$ proton source up to 5 mA to be designed and
built;

\itemnew the existing equipment to be removed from three definite straight
sections (4.87 m each);

\itemnew three partial siberian snakes  to be installed in the U70
main ring in these
three cleaned straight sections;

\itemnew the correction of the U70 vertical orbit to be done with
+/-5 mm accuracy;

\itemnew an absolute(polarized jet target) and a relative polarimetry to be
instrumented and installed into the U70 environment.

\end{itemize}

Acceleration of the polarized proton beam at U70 gives a brand new
opportunity for the high energy spin physics in the new kinematic region -
at moderate p$_T$ (up to 5 GeV/c) and large x (parton momentum).
The presented spin program includes five miscellaneous sets
of measurements :

\begin{itemize}

\itemnew polarization and double spin transverse asymmetry in elastic
pp-scattering at large $p_T$;

\itemnew single spin asymmetry in inclusive charged hadron production ;

\itemnew miscellaneous spin parameters in hyperon production;

\itemnew transversity in Drell-Yan muon pairs;

\itemnew longitudinal double-spin asymmetry in charmonium production.

\end{itemize}

The results will be complementary to those which might be obtained
at COMPASS, HERMES, RHIC, JLaB, GSI and JPARC. \\

The experiments at U70 at moderate p$_T$ and large x will be the
experiments of a new generation which will allow us to significantly excel
the previous results in spin physics (in terms of statistics,
wide kinematic region, types of hadron reactions). We hope that the
results of spin studies at U70 will allow us to reject a part
of proposed theoretical models and emphasize the most probable
mechanisms of origion of spin effects.\\

Finally, we should mention that spin measurements at U-70 with
polarized proton beam would probe the fundamental couplings
of the underlying Lagrangian
and investigate the spin structure of the nucleon.
A variety  of one- and two-spin asymmetries could be
measured. As it has often happened in the past,
these spin measurements might bring unexpected new results;
this would certainly stimulate the development of new
theoretical ideas.\\

\end{document}